# Zero-Curvature Formalism of Supersymmetric Principal Chiral Model

U. Saleem[1] and M. Hassan[2]

*Department of Physics,*
*University of the Punjab,*
*Quaid-e-Azam Campus,*
*Lahore-54590, Pakistan.*

**Abstract**

We investigate one-parameter family of transformation on superfields of super principal chiral model and obtain different zero-curvature representations of the model. The parametric transformation is related to the super Riccati equations and an infinite set of local and non-local conservation laws is derived. A Lax representation of the model is presented which gives rise to a superspace monodromy operator.

---

[1]usman_physics@yahoo.com
[2]mhassan@physics.pu.edu.pk

# 1 Introduction

There has been much interest in the study of classical as well as quantum integrability of non-linear sigma models for the last three decades [1]-[14]. Pohlmeyer and Lüscher investigated one-parameter family of transformations which is responsible for the existence of a linear system associated with the non-linear field equations of sigma model as a result of which a parametric Bäcklund transformation (BT) and an infinite set of local as well as non-local conservations laws is obtained [6, 7]. The local and non-local conservation laws were further investigated by different authors [8]-[28]. The local conserved quantities and their involution of bosonic and supersymmetric principal chiral model (SPCM) has been investigated quite recently [25, 26].

In this paper, we investigate a one-parameter family of transformation on superfields of field equations of super principal chiral model leading to a superfield Lax formalism of the model. The associated linear system in superspace is obtained, which is then expressed as a superspace zero-curvature condition. The linear system is then related to the super Bäcklund transformation and super Riccati equations of SPCM. A fermionic and bosonic Lax type representation is obtained which is associated with the monodromy operator of SPCM. The linear system in superspace is then used to generate bosonic non-local superfield currents of SPCM, which, when expressed in terms of components, coincide with those obtained in [19, 20, 26].

# 2 Super Principal Chiral Model

The superspace lagrangian of super principal chiral model is

$$\mathcal{L} = \frac{1}{2}\text{Tr}(D_+ G^{-1} D_- G),$$

where $D_\pm = \frac{\partial}{\partial \theta^\pm} - i\theta^\pm \partial_\pm$ are superspace covariant derivatives and the $G(x^\pm, \theta^\pm)$ are matrix superfields taking values in a Lie group $\mathcal{G}$. The superfield $G(x^\pm, \theta^\pm)$ is function of space coordinates $x^\pm$ and anti-commuting coordinates $\theta^\pm$. The superfield constraints of the model are

$$G(x,\theta)G^{-1}(x,\theta) = 1 = G^{-1}(x,\theta)G(x,\theta).$$

The model has global chiral symmetry i.e. the superspace lagrangian $\mathcal{L}$ is invariant under

$$\mathcal{G}_L \times \mathcal{G}_R : \qquad G(x^\pm, \theta^\pm) = \mathcal{U} G(x^\pm, \theta^\pm) \mathcal{V}^{-1},$$

where $\mathcal{U}$ and $\mathcal{V}$ are $\mathcal{G}_L$ and $\mathcal{G}_R$ valued matrix superfields respectively. The Noether conserved superfield currents associated with global transformation are

$$J^L_\pm = \mathrm{i} D_\pm G G^{-1}, \qquad J^R_\pm = -\mathrm{i} G^{-1} D_\pm G,$$

where $J^{R,L}_\pm$ are Grassmann odd and are Lie algebra valued, i.e. $J_\pm = J^a_\pm T^a$, where $\{T^a\}$ is the set of generators of Lie algebra of $\mathcal{G}$. The conservation equation and zero-curvature condition for left and right superfield currents are respectively

$$D_- J_+ - D_+ J_- = 0, \qquad (2.1)$$

$$D_- J_+ + D_+ J_- + \mathrm{i}\{J_+, J_-\} = 0, \qquad (2.2)$$

where $\{\}$ is the anticommutator. The superspace equations (2.1) and (2.2) can be combined to

$$D_- J_+ = D_+ J_- = -\frac{\mathrm{i}}{2}\{J_+, J_-\}. \qquad (2.3)$$

The equation (2.3) holds both for left and right superfield currents.

To find the component content of the model, we expand each superfield $G(x^\pm, \theta^\pm)$ as

$$G(x, \theta) = g(x)(1 + \mathrm{i}\theta^+ \psi_+(x) + \mathrm{i}\theta^- \psi_-(x) + \mathrm{i}\theta^+\theta^-\ F(x)), \qquad (2.4)$$

where $\psi_\pm$ are Majorana spinors and $F(x)$ is the auxiliary fields [3]. The superfield has an algebraic equation of motion, and eliminating it the SPCM lagrangian becomes

$$\begin{aligned} L = & -\frac{1}{2}\mathrm{Tr}\Big(g^{-1}\partial_+ g g^{-1}\partial_- g + \mathrm{i}\psi_+(\partial_- \psi_+ + \frac{1}{2}[g^{-1}\partial_- g, \psi_+]) \\ & + \frac{\mathrm{i}}{2}\psi_-(\partial_+ \psi_- + \frac{1}{2}[g^{-1}\partial_+ g, \psi_-]) + \frac{1}{2}\psi_+^2 \psi_-^2\Big). \end{aligned} \qquad (2.5)$$

The equation of motion for SPCM can be found directly from equation (2.5) using Euler-Lagrange equations. The component expansion of any superfield current of SPCM can be written as

$$J_\pm = \psi_\pm + \theta^\pm j_\pm - \frac{1}{2}\mathrm{i}\theta^\mp \{\psi_+, \psi_-\} - \mathrm{i}\theta^+\theta^-\left(\partial_\pm \psi_\mp - [j_\pm, \psi_\mp] - \frac{\mathrm{i}}{2}[\psi_\pm^2, \psi_\mp]\right),$$

---

[3] Our notation conventions are as follows. The two-dimensional Minkowski metric is $\eta_{\mu\nu} = \mathrm{diag}(+1, -1)$, the Dirac algebra $\{\gamma_\mu, \gamma_\nu\} = 2\eta_{\mu\nu}$ is satisfied by the $\gamma$-matrices

$$\gamma_0 = \begin{pmatrix} 0 & \mathrm{i} \\ -\mathrm{i} & 0 \end{pmatrix}, \ \gamma_1 = \begin{pmatrix} 0 & \mathrm{i} \\ \mathrm{i} & 0 \end{pmatrix}.$$

The Dirac spinor has two components $\psi_\pm$ called chiral spinors, and we shall assume that $\psi_\pm$ are real (Majorana) i.e. $\psi^*_\pm = \psi_\pm$. Under a Lorentz transformation $x^\pm, \partial_\pm$ and $\psi_\pm$ transform as $x^\pm \longmapsto e^{\mp\Lambda}x^\pm, \partial_\pm \longmapsto e^{\mp\Lambda}\partial_\pm$ and $\psi_\pm \to e^{\mp\frac{1}{2}\Lambda}\psi_\pm$ where $\Lambda$ is the rapidity of the Lorentz boost. The rule for raising and lowering spinor indices is $\psi^\pm = \pm\psi_\mp$.

where $j_\pm = -(g^{-1}\partial_\pm g + \mathrm{i}\,\psi_\pm^2)$. Substituting these into the superspace equations of motion, collecting terms and writing $h_\pm = \psi_\pm^2$, we get equations of motion for fermionic and bosonic fields of SPCM

$$\partial_\pm \psi_\mp - \frac{1}{2}[j_\pm, \psi_\mp] - \frac{\mathrm{i}}{4}[h_\pm, \psi_\mp] = 0, \qquad \partial_- j_+ + \partial_+ j_- = 0,$$

along with

$$\partial_\mp j_\pm = -\frac{1}{2}[j_\pm, j_\mp] + \frac{\mathrm{i}}{4}[j_\mp, h_\pm] - \frac{\mathrm{i}}{4}[j_\pm, h_\mp] + \frac{1}{4}[h_\pm, h_\mp].$$

We use fermion equations of motion to get the following equations:

$$\partial_- j_+ - \partial_+ j_- + [j_+, j_-] = \mathrm{i}\partial_- h_+ - \mathrm{i}\partial_+ h_-, \qquad \partial_\mp (\mathrm{i} h_\pm) = -\frac{1}{2}[\mathrm{i} h_\pm, j_\mp + \frac{\mathrm{i}}{2} h_\mp].$$

## 3 Superfield Lax Formalism

The superfield Lax representation of SPCM can be obtained by defining a one-parameter family of transformation on the superfields of SPCM. A one-parameter family of transformations on the superfields $G(x,\theta)$ is defined by a matrix superfields $\mathcal{U}^{(\gamma)}$ and $\mathcal{V}^{(\gamma)}$ where $\gamma$ is a real number. The transformation on the superfield $G(x^\pm, \theta^\pm)$ is

$$G(x^\pm, \theta^\pm) \mapsto G^{(\gamma)}(x^\pm, \theta^\pm) = \mathcal{U}^{(\gamma)} G(x^\pm, \theta^\pm) \mathcal{V}^{(\gamma)-1}, \tag{3.1}$$

where $\mathcal{U}^{(\gamma)}$ and $\mathcal{V}^{(\gamma)}$ belong to $\mathcal{G}$. Here we choose the boundary values $\mathcal{V}^{(1)} = 1$, $\mathcal{U}^{(1)} = 1$ or $G^{(1)} = G$. It can be checked that if $G$ is any classical solution, so is $G^{(\gamma)}$, provided that

$$D_\pm \mathcal{U}^{(\gamma)} = -\frac{\mathrm{i}}{2}(1 - \gamma^{\mp 1}) J_\pm^L \mathcal{U}^{(\gamma)}, \qquad D_\pm \mathcal{V}^{(\gamma)} = -\frac{\mathrm{i}}{2}(1 - \gamma^{\mp 1}) J_\pm^R \mathcal{V}^{(\gamma)}. \tag{3.2}$$

The compatibility conditions for these equations are

$$\left\{(1-\gamma^{-1})D_- J_+^L + (1-\gamma)D_+ J_-^L + \mathrm{i}(1 - \frac{1}{2}(\gamma + \gamma^{-1}))\{J_+^L, J_-^L\}\right\} \mathcal{U}^{(\gamma)} = 0,$$

$$\left\{(1-\gamma^{-1})D_- J_+^R + (1-\gamma)D_+ J_-^R + \mathrm{i}(1 - \frac{1}{2}(\gamma + \gamma^{-1}))\{J_+^R, J_-^R\}\right\} \mathcal{V}^{(\gamma)} = 0.$$

From the transformation on the superfields $G(x^\pm, \theta^\pm)$, one can easily find the transformation on Lie algebra valued Noether superfield current:

$$J_\pm^L \mapsto J_\pm^{L(\gamma)} = \gamma^{\mp 1} \mathcal{U}^{(\gamma)-1} J_\pm^L \mathcal{U}^{(\gamma)}, \qquad J_\pm^R \mapsto J_\pm^{R(\gamma)} = \gamma^{\mp 1} \mathcal{V}^{(\gamma)-1} J_\pm^R \mathcal{V}^{(\gamma)}.$$

Since the one-parameter family of transformation maps classical solutions to new classical solutions, these superfield currents are conserved in superspace for any value of $\gamma$:

$D_+ J_-^{L,R(\gamma)} - D_- J_+^{L,R(\gamma)} = 0$. From now and onwards, we shall only use $J_\pm^R$ and drop the superscript to write $J_\pm = J_\pm^R$. The associated linear system of SPCM is then written as

$$D_\pm \mathcal{V}(t,x,\theta;\lambda) = \mathcal{A}_\pm^{(\lambda)} \mathcal{V}(t,x,\theta;\lambda), \tag{3.3}$$

where the odd superfields $\mathcal{A}_\pm^{(\lambda)}$ are given by

$$\mathcal{A}_\pm^{(\lambda)} = \pm \frac{i\lambda}{1 \mp \lambda} J_\pm. \tag{3.4}$$

The parameter $\lambda$ is the spectral parameter and is related to parameter $\gamma$ by $\lambda = \frac{1-\gamma}{1+\gamma}$. The compatibility condition of the linear system (3.3) reduces to a fermionic zero-curvature condition for odd superfields $\mathcal{A}_\pm^{(\lambda)}$ as follows:

$$\{D_+ - \mathcal{A}_+^{(\lambda)}, D_- - \mathcal{A}_-^{(\lambda)}\} \equiv D_+ \mathcal{A}_-^{(\lambda)} + D_- \mathcal{A}_+^{(\lambda)} - \{\mathcal{A}_+^{(\lambda)}, \mathcal{A}_-^{(\lambda)}\} = 0. \tag{3.5}$$

The superspace Grassmann odd operators $\mathcal{L}_\pm^{(\lambda)} = D_\pm - \mathcal{A}_\pm^{(\lambda)}$, satisfy Lax type equations in superspace $D_\mp \mathcal{L}_\pm^{(\lambda)} = \{\mathcal{A}_\mp^{(\lambda)}, \mathcal{L}_\pm^{(\lambda)}\}$. By applying $D_\pm$ on (3.3) respectively, one gets a linear system in terms of even superfields $\tilde{\mathcal{A}}_\pm^{(\lambda)}$:

$$\partial_\pm \mathcal{V}(t,x,\theta;\lambda) = \tilde{\mathcal{A}}_\pm^{(\lambda)} \mathcal{V}(t,x,\theta;\lambda), \tag{3.6}$$

where the even superfields $\tilde{\mathcal{A}}_\pm^{(\lambda)}$ are

$$\tilde{\mathcal{A}}_\pm^{(\lambda)} = \left\{ \mp \left(\frac{\lambda}{1\mp\lambda}\right) D_\pm J_\pm + i \left(\frac{\lambda}{1\mp\lambda}\right)^2 J_\pm^2 \right\}. \tag{3.7}$$

The compatibility condition of equation (3.7) now reduces to a bosonic zero-curvature condition for even superfields $\tilde{\mathcal{A}}_\pm^{(\lambda)}$:

$$[\partial_+ - \tilde{\mathcal{A}}_+^{(\lambda)}, \partial_- - \tilde{\mathcal{A}}_-^{(\lambda)}] \equiv \partial_- \tilde{\mathcal{A}}_+^{(\lambda)} - \partial_+ \tilde{\mathcal{A}}_-^{(\lambda)} + [\tilde{\mathcal{A}}_+^{(\lambda)}, \tilde{\mathcal{A}}_-^{(\lambda)}] = 0. \tag{3.8}$$

The associated linear system (3.6) can be re-expressed in term of space - time coordinates:

$$\partial_0 \mathcal{V}(t,x,\theta;\lambda) = \tilde{\mathcal{A}}_0^{(\lambda)} \mathcal{V}(t,x,\theta;\lambda), \qquad \partial_1 \mathcal{V}(t,x,\theta;\lambda) = \tilde{\mathcal{A}}_1^{(\lambda)} \mathcal{V}(t,x,\theta;\lambda), \tag{3.9}$$

with the superfields $\tilde{\mathcal{A}}_0^{(\lambda)}$ and $\tilde{\mathcal{A}}_1^{(\lambda)}$ are defined by

$$\tilde{\mathcal{A}}_0^{(\lambda)} = \frac{1}{2} \left\{ \left(\frac{-\lambda}{1-\lambda}\right) D_+ J_+ + \left(\frac{\lambda}{1+\lambda}\right) D_- J_- + i\left(\frac{\lambda}{1-\lambda}\right)^2 J_+^2 + i\left(\frac{\lambda}{1+\lambda}\right)^2 J_-^2 \right\},$$

$$\tilde{\mathcal{A}}_1^{(\lambda)} = \frac{1}{2} \left\{ \left(\frac{-\lambda}{1-\lambda}\right) D_+ J_+ - \left(\frac{\lambda}{1+\lambda}\right) D_- J_- + i\left(\frac{\lambda}{1-\lambda}\right)^2 J_+^2 - i\left(\frac{\lambda}{1+\lambda}\right)^2 J_-^2 \right\}.$$

The compatibility condition of the system (3.9) is

$$[\partial_0 - \tilde{\mathcal{A}}_0^{(\lambda)}, \partial_1 - \tilde{\mathcal{A}}_1^{(\lambda)}] \equiv \partial_0 \tilde{\mathcal{A}}_1^{(\lambda)} - \partial_0 \tilde{\mathcal{A}}_1^{(\lambda)} + \left[\tilde{\mathcal{A}}_0^{(\lambda)}, \tilde{\mathcal{A}}_1^{(\lambda)}\right] = 0. \tag{3.10}$$

This is essentially the zero-curvature condition which leads to a Lax type representation of SPCM. We can now define Grassmann even operators in superspace: $\tilde{\mathcal{L}}_1^{(\lambda)} = \partial_1 - \tilde{\mathcal{A}}_1^{(\lambda)}$; $\tilde{\mathcal{L}}_0^{(\lambda)} = \partial_0 - \tilde{\mathcal{A}}_0^{(\lambda)}$, obeying the following equations:

$$\partial_0 \tilde{\mathcal{L}}_1^{(\lambda)} = \left[\tilde{\mathcal{A}}_0^{(\lambda)}, \tilde{\mathcal{L}}_1^{(\lambda)}\right], \qquad \partial_1 \tilde{\mathcal{L}}_0^{(\lambda)} = \left[\tilde{\mathcal{A}}_1^{(\lambda)}, \tilde{\mathcal{L}}_0^{(\lambda)}\right].$$

We define the theory on the spatial interval $[-a, a]$ and the superfields $\tilde{\mathcal{A}}_0^{(\lambda)}$ and $\tilde{\mathcal{A}}_1^{(\lambda)}$ are subject to boundary conditions: $\tilde{\mathcal{A}}_0^{(\lambda)}(a) = \tilde{\mathcal{A}}_0^{(\lambda)}(-a)$, $\tilde{\mathcal{A}}_1^{(\lambda)}(a) = \tilde{\mathcal{A}}_1^{(\lambda)}(-a)$. The equation satisfied by superspace monodromy operator $T_\lambda(x, \theta)$ is

$$\frac{\partial}{\partial x} T_\lambda(x, \theta) = \tilde{\mathcal{A}}_1^{(\lambda)} T_\lambda(x, \theta), \tag{3.11}$$

with boundary condition $T_\lambda(-a) = 1$. The solution of equation(3.11) is

$$T_\lambda(x, \theta) = P \exp\left(-\int_{-a}^{x} dy\, \tilde{\mathcal{A}}_1^{(\lambda)}(y, \theta; \lambda)\right), \tag{3.12}$$

where $P$ is the path-ordered operator. The operator $T_\lambda(x, \theta)$ obeys

$$\partial_0 T_\lambda(x, \theta) = \left[\tilde{\mathcal{A}}_0^{(\lambda)}(a), T_\lambda(a)\right],$$

which the Lax form of monodromy operator in superspace. This can be used to generate an infinite sequence of local and non-local conservation laws as detailed in section 4.

## 4 Superfield Conserved Quantities

### 4.1 Local Conserved Quantities

From the superspace equations of motion (2.3) of SPCM, it is straightforward to derive the following set of local conservation laws [26]:

$$D_\pm \text{Tr}(J_\mp^{2n+1}) = 0, \quad D_\pm \text{Tr}(J_\mp^{2n-1} J_{\mp\mp}) = 0, \quad \text{with} \quad J_{\mp\mp} = D_\mp J_\mp + \mathrm{i} J_\mp^2, \tag{4.1}$$

where the values of $n$ are precisely the exponents of the Lie algebra of $\mathcal{G}$.

The local conserved quantities of SPCM also arise from the linear system (3.3) via super Bäcklund transformation (SBT) or equivalently super Riccati equations. The linear system (3.3) reduces to the following set of super Bäcklund transformation (SBT) [21, 22]:

$$\pm D_\pm \left(G^{-1}\bar{G}\right) = \mathrm{i} J_\pm - \mathrm{i}\bar{J}_\pm, \tag{4.2}$$

with the constraint $G^{-1}\bar{G}+\bar{G}^{-1}G = 2\lambda^{-1}I$, where $G$ and $\bar{G}$ are solutions of superfield equations. The SBT (4.2) can be recast into the following set of compatible Riccati equations in superspace:

$$D_\pm \mathcal{N}(\lambda) = \frac{\mathrm{i}\lambda}{2(1\mp\lambda)}\left(J_\pm + \mathcal{N}(\lambda)J_\pm\mathcal{N}(\lambda) - 2\lambda^{-1}J_\pm\mathcal{N} \mp [\mathcal{N}(\lambda), J_\pm]\right), \tag{4.3}$$

where $\mathcal{N} = G^{-1}\bar{G}$ is an even matrix superfield. We can linearize equations (4.3) by taking $\mathcal{N} = \mathcal{V}_1\mathcal{V}_2^{-1}$, as follows:

$$D_\pm \mathcal{V} = \frac{\mathrm{i}J_\pm}{2(1\mp\lambda)}\begin{pmatrix}\pm\lambda-2 & \lambda \\ -\lambda & \pm\lambda\end{pmatrix}\mathcal{V}, \quad \text{where} \quad \mathcal{V} = \begin{pmatrix}\mathcal{V}_1 \\ \mathcal{V}_2\end{pmatrix}. \tag{4.4}$$

The diagonalization of (4.4) gives (3.3). By using equations (2.1), (2.2), (4.3), we can derive a series of conservation laws:

$$(1+\lambda)D_-\operatorname{Tr}(\mathcal{N}(\lambda)J_+) + (1-\lambda)\,D_+\operatorname{Tr}(\mathcal{N}(\lambda)J_-) = 0. \tag{4.5}$$

Expanding $\mathcal{N}(\lambda)$ as a power series in $\lambda$: $\mathcal{N}(\lambda) = \sum_{k=0}^\infty \lambda^k \mathcal{N}_k$, one can generate $\lambda$-independent conservation laws. The explicit determination of $\mathcal{N}_k$ requires the solution of a system of algebraic matrix equations obtained recursively. This system is not easier to solve so that one could get explicit form of local conservation laws.

## 4.2  Non-local Conserved Quantities

The non-local conserved quantities of SPCM can be extracted directly from the superfield $\mathcal{V}^{(\gamma)}$. We assume spatial boundary conditions such that the superfield $J_\pm$ vanishes as $x \to \pm\infty$. Then (3.9) implies that $\mathcal{V}(t, \pm\infty, \theta; \lambda)$ are independent of time. The residual freedom in the solution for $\mathcal{V}^{(\gamma)}$ allows us to fix $\mathcal{V}(t, -\infty, \theta; \lambda)$ equal to the unit matrix. We are then left with a time independent function: $\mathcal{Q}(\lambda) = \mathcal{V}(t, \infty; \lambda)$. Expanding $\mathcal{Q}(\lambda)$ as a power series in $\lambda$ gives infinitely many conserved quantities

$$\mathcal{Q}(\lambda) = \sum_{k=0}^\infty \lambda^k \mathcal{Q}^{(k)}, \qquad \frac{d\mathcal{Q}^{(k)}}{dt} = 0.$$

In order to derive explicit expressions for these conserved quantities in terms of superfields, we write equation (3.9) as

$$\mathcal{V}(t,x,\theta;\lambda) = 1 + \frac{1}{2}\int_{-\infty}^{x} dy \left\{\left(\frac{-\lambda}{1-\lambda}\right) D_+ J_+ - \left(\frac{\lambda}{1+\lambda}\right) D_- J_- + \right.$$
$$\left. + \mathrm{i}\left(\frac{\lambda}{1-\lambda}\right)^2 J_+^2 - \mathrm{i}\left(\frac{\lambda}{1+\lambda}\right)^2 J_-^2 \right\} \mathcal{V}(t,y,\theta;\lambda). \quad (4.6)$$

When we expand the superfield $\mathcal{V}(t,x,\theta;\lambda)$ as a power series in $\lambda$,

$$\mathcal{V}(t,x,\theta;\lambda) = \sum_{k=0}^{\infty} \lambda^k \mathcal{X}_k(t,x,\theta), \quad (4.7)$$

and compare the coefficients of powers of $\lambda$, one gets a series of conserved non-local superfield currents, which upon integration give non-local conserved quantities. The expressions for the first few cases are

$$\mathcal{Q}^{(1)a} = -\frac{1}{2}\int_{-\infty}^{\infty} dy \, (D_+ J_+^a + D_- J_-^a)(t,y,\theta),$$
$$\mathcal{Q}^{(2)a} = \int_{-\infty}^{\infty} dy \left(-\frac{1}{2}(D_+ J_+^a - D_- J_-^a)(t,y,\theta) + \frac{\mathrm{i}}{4} f^{abc}(J_+^b J_+^c - J_-^b J_-^c)(t,y,\theta)\right.$$
$$\left. + \frac{1}{8} f^{abc}(D_+ J_+^b + D_- J_-^b)(t,y,\theta) \int_{-\infty}^{y} dz \, (D_+ J_+^c + D_- J_-^c)(t,z,\theta)\right). \quad (4.8)$$

These conserved quantities are exactly the same as obtained in [26] using iterative method. By substituting the expansion (4.7) in (3.3) we have

$$D_\pm \sum_{k=0}^{\infty} \lambda^k \mathcal{X}^{(k)} = \pm \mathcal{D}_\pm \sum_{k=0}^{\infty} \lambda^k \mathcal{X}^{(k)}.$$

The covariant derivatives $\mathcal{D}_\pm$ is defined as

$$\mathcal{D}_\pm \mathcal{X}^{(k)} = D_\pm \mathcal{X}^{(k)} + \mathrm{i}[J_\pm, \mathcal{X}^{(k)}] \quad \Rightarrow \quad \{\mathcal{D}_+, \mathcal{D}_-\} = 0.$$

We define superfield currents $J_\pm^{(k)}$ for $k = 0, 1, \ldots$ which are conserved in superspace such that

$$D_- J_+^{(k)} - D_+ J_-^{(k)} = 0, \quad \Leftrightarrow \quad J_\pm^{(k)} = \pm \mathrm{i} D_\pm \mathcal{X}^{(k)}.$$

An infinite sequence of conserved non-local superfield currents can be obtained by iteration [26]:

$$J_\pm^{(k+1)} = \mathcal{D}_\pm \mathcal{X}^{(k)}, \quad \Rightarrow \quad D_- J_+^{(k+1)} - D_+ J_-^{(k+1)} = 0.$$

This establishes the equivalence of superfield Lax formalism and iterative construction of conserved non-local superfield currents.

# 5 Lax Formalism on Component Fields of SPCM and Conserved Quantities

The transformation (3.1) is equivalent to the following set of transformation on component fields of SPCM:

$$g \to g^{(\gamma)} = U^{(\gamma)} g V^{(\gamma)-1}, \quad \psi_{\pm}^R \to \psi_{\pm}^{R(\gamma)} = V^{(\gamma)} \psi_{\pm}^R V^{(\gamma)-1}, \quad \psi_{\pm}^L \to \psi_{\pm}^{L(\gamma)} = U^{(\gamma)} \psi_{\pm}^L U^{(\gamma)-1}.$$

The conserved currents $j_{\pm}$ transforms as: $j_{\pm} \to j_{\pm}^{(\gamma)} = \gamma^{\mp 1} V^{(\gamma)-1} j_{\pm} V^{(\gamma)}$, where $V^{(\gamma)}$ and $U^{(\gamma)}$ are the leading bosonic components of the matrix superfields $\mathcal{V}^{(\gamma)}$ and $\mathcal{U}^{(\gamma)}$. The associated bosonic linear system in component fields is written as

$$\partial_{\pm} V(t, x; \lambda) = A_{\pm}^{(\lambda)} V(t, x; \lambda),$$

where $A_{\pm}^{(\lambda)}$ is defined as

$$A_{\pm}^{(\lambda)} = \left\{ \mp \left( \frac{\lambda}{1 \mp \lambda} \right) j_{\pm} + i \left( \frac{\lambda}{1 \mp \lambda} \right)^2 h_{\pm} \right\}.$$

The compatibility condition of the linear system is the zero-curvature condition,

$$[\partial_+ - A_+^{(\lambda)}, \partial_- - A_-^{(\lambda)}] \equiv \partial_- A_+^{(\lambda)} - \partial_+ A_-^{(\lambda)} + [A_+^{(\lambda)}, A_-^{(\lambda)}] = 0.$$

The operators $L_{\pm}^{(\lambda)} = \partial_{\pm} - A_{\pm}^{(\lambda)}$ obey the Lax equations: $\partial_{\mp} L_{\pm}^{(\lambda)} = \left[ A_{\mp}^{(\lambda)}, L_{\pm}^{(\lambda)} \right]$. In terms of space time coordinates, the associated linear system can be expressed as

$$\partial_0 V(t, x; \lambda) = A_0^{(\lambda)} V(t, x; \lambda), \qquad \partial_1 V(t, x; \lambda) = A_1^{(\lambda)} V(t, x; \lambda), \tag{5.1}$$

with

$$A_0^{(\lambda)} = -\frac{\lambda}{1-\lambda^2} \left\{ j_1 + \lambda j_0 - \frac{i}{2}\lambda \left( \frac{1+\lambda}{1-\lambda} \right) h_+ - \frac{i}{2}\lambda \left( \frac{1-\lambda}{1+\lambda} \right) h_- \right\},$$

$$A_1^{(\lambda)} = -\frac{\lambda}{1-\lambda^2} \left\{ j_0 + \lambda j_1 - \frac{i}{2}\lambda \left( \frac{1+\lambda}{1-\lambda} \right) h_+ + \frac{i}{2}\lambda \left( \frac{1-\lambda}{1+\lambda} \right) h_- \right\}.$$

The operators $L_1^{(\lambda)} = \partial_1 - A_1^{(\lambda)}$; $L_0^{(\lambda)} = \partial_0 - A_0^{(\lambda)}$, obey the following equations:

$$\partial_0 L_1^{(\lambda)} = \left[ A_0^{(\lambda)}, L_1^{(\lambda)} \right], \qquad \partial_1 L_0^{(\lambda)} = \left[ A_1^{(\lambda)}, L_0^{(\lambda)} \right].$$

To find the component content of the Bäcklund transformation, we expand each superfield as in equation (2.4). The set of Bäcklund transformation on component field is

$$\pm \partial_{\pm} \left( g^{-1} \bar{g} \right) = \bar{j}_{\pm} - j_{\pm}, \tag{5.2}$$

with constraint $g^{-1}\bar{g} + \bar{g}^{-1}g = 2\lambda^{-1} I$. The matrix superfield $\mathcal{N}(\lambda)$ can also be expanded in term of components as: $\mathcal{N} = N + i\theta^+ N_+ + i\theta^- N_- + i\theta^+\theta^- H$.

Substituting expansion of $\mathcal{N}$ and $\mathcal{V}$ in equations (4.3), we get

$$\partial_\pm N(\lambda) = \frac{\lambda}{2(1 \mp \lambda)}\left[\left(-j_\pm - N(\lambda)j_\pm N(\lambda) + 2\lambda^{-1}j_\pm N(\lambda) \pm [N(\lambda), j_\pm]\right) + (iN(\lambda)\psi_\pm N_\pm(\lambda) - \right.$$
$$\left. - iN_\pm(\lambda)\psi_\pm N(\lambda) - 2i\lambda^{-1}\psi_\pm N_\pm(\lambda) \pm i\{N_\pm(\lambda), \psi_\pm\})\right], \qquad (5.3)$$

with

$$N_\pm(\lambda) = \frac{\lambda}{2(1 \mp \lambda)}\left(\psi_\pm + N(\lambda)\psi_\pm N(\lambda) - 2\lambda^{-1}\psi_\pm N(\lambda) \mp [N(\lambda), \psi_\pm]\right).$$

The Riccati equation (5.3) can be linearized by taking $N = V_1 V_2^{-1}$:

$$\partial_\pm \begin{pmatrix} V_1 \\ V_2 \end{pmatrix} = \frac{1}{2(1 \mp \lambda)}\left\{j_\pm \begin{pmatrix} 2 \mp \lambda & -\lambda \\ \lambda & \mp \lambda \end{pmatrix} + ih_\pm \begin{pmatrix} \mp 2 & -\lambda \\ \lambda & 0 \end{pmatrix}\right\}\begin{pmatrix} V_1 \\ V_2 \end{pmatrix}. \qquad (5.4)$$

The diagonalization of the matrices is equivalent to the well-known linear system of SPCM. From the Riccati equations (5.3) the following conservation equation directly follows:

$$(1 + \lambda)\partial_- \text{Tr}\,(N(\lambda)j_+ + iN_+(\lambda)\psi_+) - (1 - \lambda)\partial_+ \text{Tr}\,(N(\lambda)j_- + iN_-(\lambda)\psi_-) = 0. \qquad (5.5)$$

The expansion of $N$ in $\lambda$ would yield the local conservation laws of the model. Once the explicit form of conservation laws (5.5) is obtained, one should be able to relate them to the local conservation laws of (4.1).

From the linear system (5.1), one can easily find conserved quantities

$$Q^{(1)a} = -\int_{-\infty}^{\infty} dy\, j_0^a(t, y),$$
$$Q^{(2)a} = \int_{-\infty}^{\infty} dy\, \left(-j_1^a(t, y) + \frac{i}{2}(h_+^a - h_-^a)(t, y)\right.$$
$$\left. + \frac{1}{2}f^{abc} j_0^b(t, y) \int_{-\infty}^{y} dz\, j_0^c(t, z)\right),$$

which are the non-local conserved quantities obtained in [18]-[20], [26][4].

In summary: we have investigated one-parameter family of transformation on superfields of SPCM leading to a superfield Lax formalism. The linear system associated to non-linear superfield equations is obtained and as a result bosonic and fermionic zero-curvature conditions appear. The superfield Lax formalism is then used to generate superfield local conserved quantities of SPCM. The superfield Lax formalism is shown to be equivalent to the iteration construction of super non-local conserved quantities. The linear system is then related to super Bäcklund transformation (SBT) and super Riccati equations. The linear system in superspace is further analyzed to obtain expressions of non-local superfield currents. At the end, we applied these considerations to the component content of SPCM. There are many interesting directions in which the work can be extended. One way is to obtain the Poisson bracket algebra of monodromy operator of SPCM and then to investigate the algebra of local conserved quantities of SPCM. It would be interesting to relate superfield local conserved quantities (4.1) with (4.5). The Lax representation of SPCM can further be investigated to incorporate inverse scattering method. The zero-curvature structure and Lax formalism of SPCM can also be investigated on non-commutative spaces.

---
[4]These conserved quantities generate a non-local symmetry (Yangian) as in the bosonic case. There are two copies of Yangians corresponding to $L$ and $R$ currents [25]-[28].

## Acknowledgment

We acknowledge the enabling role of the Higher Education Commission Pakistan and appreciate its financial support through "Merit Scholarship Scheme for PhD studies in Science & Technology (200 Scholarships)". We also acknowledge CERN scientific information Service (publication requests).

## References

[1] O. Babelon, D. Bernand, M. Talon, *Introduction to classical integrable systems*, Cambridge University Press Cambridge UK (2003).

[2] E. Abdalla, M. C. B. Abdalla and K. Rothe, *Non-perturbative methods in two dimensional quantum field theory*, World Scientific Singapore (1991).

[3] G. L. Lamb Jr, *Elements of soliton theory*, John Wiley & Sons New York (1980).

[4] A. Das, *Integrable models*, World Scientific Singapore (1989).

[5] E. Abdalla, M. C. B. Abdalla, Phys. Lett. **B152** (1985) 59.

[6] K. Pohlmeyer, Comm. Math. Phys. **46** (1976) 2017.

[7] M. Lüscher and K. Pohlmeyer, Nucl. Phys. **B137** (1978) 46.

[8] E. Brézin, C. Itzykson, J. Zinn-Justin and J.-B. Zuber, Phys. Lett. **B 82** (1979) 442.

[9] Y-S Wu , Nucl. Phys. **B211** (1983) 160.

[10] A. M. Polyakov, Phys. Lett. **B 82** (1979) 247.

[11] A. T. Ogielski, Phys. Rev. **D21** (1980) 406.

[12] H. J. de Vega, Phys. Lett. **B 87** (1979) 233.

[13] D. Bernard, Phys. Lett. **B 279** (1992) 78.

[14] N. J. MacKay, Phys. Lett. **B 281** (1992) 90.

[15] D. Levi, O. Ragnisco, and A. Sym. Lett. Nuovo Cimento **33** (1982) 401.

[16] K. Scheler, Z. Phys. **C6** (1980) 365.

[17] D. Levi, O. Ragnisco, Phys. Lett. **A87** (1982) 381.

[18] E. Corrigan and C. K. Zachos, Phys. Lett. **B88** (1979) 273.

[19] T. L. Curtright, Phys. Lett. **B88** (1979) 276.

[20] T. L. Curtright and C. K. Zachos, Phys. Rev. **D21** (1980) 411.

[21] Z. Popowicz and L. L. Chau Wang, Phys. Lett. **B98** (1981) 253.

[22] L. L. Chau and H. C. Yen, Phys. Lett. **B177** (1986) 368.

[23] E. Abdalla and M. Forger, Commun. Math. Phys. **104** (1986) 123.

[24] J. Barcelos-Neto, A. Das and J. Maharana, Z. Phys. **C 30** (1986) 401.

[25] J. M. Evans, M. Hassan, N. J. MacKay and A. J. Mountain, Nucl. Phys. **B561** (1999) 385.

[26] J. M. Evans, M. Hassan, N. J. MacKay and A. J. Mountain, Nucl. Phys. **B580** (2000) 605 .

[27] D. Bernard, Commun. Math. Phys. **137** (1991) 191.

[28] T. L. Curtright and C. K. Zachos, Nucl. Phys. **B402** (1993) 604.